# DISCRETE-ELEMENT SIMULATION OF PARTICLE BREAKAGE INSIDE BALL MILLS: A 2D MODEL

Luisa Fernanda Orozco[1,3], Duc-Hanh Nguyen[4], Jean-Yves Delenne[2], Philippe Sornay[3], Farhang Radjai[1]

1 Laboratoire de Mécanique et Génie Civil (LMGC), Université de Montpellier, CNRS, Montpellier, France
2 INRA, UMR IATE Montpellier, France
3 CEA, DEN, DEC, SA3E, LCU, 13108 Saint Paul les Durance, France
4 National University of Civil Engineering, Hanoi, Vietnam
luisa.orozco@umontpellier.fr

## ABSTRACT

In this paper, the grinding of powder inside a ball mill is studied using the Bonded Cell Method (BCM) implemented in the framework of Contact Dynamics. In BCM the parent particles are divided into cells that are glued to one another until, under the action of external loading, both a cohesion threshold and a certain distance threshold are reached. Numerical simulations of a rotating hollow cylinder filled with a mixture of heavy balls and powder crushable particles were carried out. Systems with balls of different sizes and/or numbers are compared in terms of the evolution of the powder particle size and specific surface. We find that, in general, the milling process is increasingly faster as the ball size increases. But energy dissipation due to increased collisions between balls slows down the grinding process and makes it energetically less efficient. On the other hand, when the total volume of balls is kept constant, the ball size is not relevant for the evolution of particle breakage except in the limit cases of very small and very large ball sizes.

## 1. INTRODUCTION

Ball mills are widely used in multiple different industries such as agronomy, mining, pharmaceutical. In these applications, ball mills are mainly used for grinding, breakage and mixing. It has been estimated that in Australia, just in the mining sector, the grinding process consumes 36% of the total energy, which corresponds to 1.3% of Australia's energy consumption [1]. For these reasons, understanding the behaviour of granular materials in ball mills is crucial for the improvement of the operational conditions and consequently a reduction on energy consumption.

Some studies have been performed in order to evaluate the performance of ball mills regarding the choice of operational parameters, material properties and milling conditions. Commonly, properties such as the particle size distribution, powder specific surface, powder density, breakage rate, collision energy and collision frequency are compared among different systems in order to evaluate the grinding energy efficiency and the effectivity of particle size reduction [2,3,4]. In many experimental studies of ball milling, the amount/range of the parameters tested is limited, and therefore inconclusive results are found. This gap may be filled by numerical simulation, which currently has its own challenge of reconciling numerical performance with the realism of the underlying physical model.

A few DEM models have been developed in order to simulate particle breakage inside drum mills. Some use the bonded particle method (BPM) in which the parent particle is composed of smaller spheres agglomerated [5]. Inside the drum, the agglomerates are ground due to the interactions with bigger unbreakable balls, walls and other agglomerates. However, in this method, the particle volume is not conserved. In another approach sometimes used, a particle is replaced by a collection of smaller spheres once a breakage criterion is achieved [6,7]. Since the particle replacement is done using a population balance model, the volume is conserved but, all the particles and their fragments keep their spherical shape, and therefore the milling process cannot be modelled in full by this methodology.

In this paper, we consider the Bonded Cell Method (BCM) in 2D, which has the advantage of keeping the volume of particles and allowing for arbitrary irregular polygonal shapes of the particles and their fragments during the milling process. We briefly introduce the BCM in the framework of the Contact Dynamics (CD) method. Then, we present the results of two groups of numerical simulations of grinding inside a ball mill, which we analyze to investigate the effect of operational parameters on the evolution of particle size and specific surface as a function of the number of revolutions. As the numerical efficiency of DEM simulations strongly declines with the number of particles, one crucial aspect of this work is to show that realistic simulations can be performed using our approach with a good representation of the ball-particle mixture.

## 2. NUMERICAL METHOD

### Bonded Cell Method (BCM)

In the BCM, each particle is modelled as an assembly of fundamental elements to which we will refer below as 'cells'. Thus, when a particle breaks, the fragments generated are smaller particles each composed of cells, the smallest fragment being a single cell (representing the lower bound on particle size). In order to define the cells configuration, a Voronoï tessellation is performed on each particle. The mean cell size ($d_{cell}$) is fixed so that a parent particle of surface $s$ consists of approximately $s/d_{cell}^2$ cells. A parameter $\alpha$ accounts for the cell shape heterogeneity, taking the value of 1 for very similar cell shapes, and 0 for very dissimilar cells. In previous work applying BCM it was found that setting $\alpha$ close to 1 leads to nearly crystalized cell configurations with higher mechanical strength [8]. To avoid such effects, in this work $\alpha$ is set to 0.5. The generated cells are convex polygons that are in perfect contact with their neighbours. In this way, each parent particle is perfectly tessellated without defects nor voids. Figure 1(a) displays an example of a collection of pentagonal particles partitioned into irregular cells.

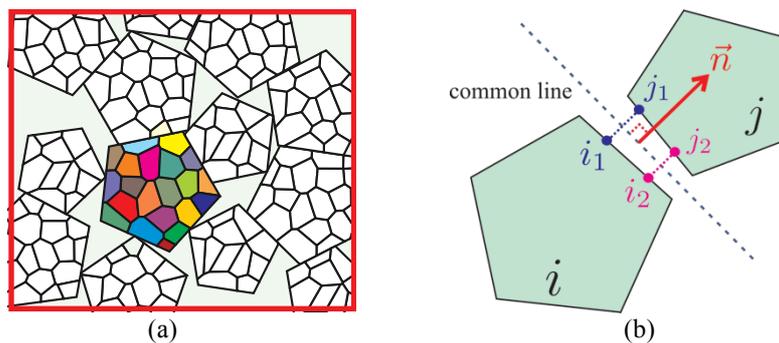

(a)          (b)

Figure 1. (a) Voronoï tessellation applied to polygonal particles. Each cell is presented in a different colour; (b) The geometry of a side-side contact between two cells.

Since the cells have a polygonal shape, various contact types can be expected, i.e. side-side, vertex-side, vertex-vertex. Initially, cohesive bonds are assigned to all side-side interfaces. The side-side contacts are represented by two points belonging to their common contact line (see Fig. 1(b)). Initially, the contact lines coincide with the common sides between cells. Mechanically, a contact can lose its cohesive status only if some strength and distance criteria are fulfilled. In our model, these criteria are defined as a function of the distance between the points of contact.

We introduce three parameters for the cohesive behaviour among cells: a normal stress threshold ($C_n$), a friction coefficient ($\mu$), and a distance threshold ($\varepsilon$). The first criterion must be fulfilled in order to allow the relative movement between the two cells to occur. In the normal direction, the stress threshold is $C_n$ and in the tangential direction is $C_t = \beta\, C_n$. The parameter $\beta$ is not a friction coefficient but simply a scalar relating normal and tangential stresses. Once two contact points reach a separation distance equal to $\varepsilon$, the cohesive status is lost irreversibly and a purely frictional behaviour will govern future interactions between the two cells with a friction coefficient $\mu$. Since the cohesion is applied at the side-side interfaces, the threshold forces are calculated as the product of the stress threshold and the interface length $\ell$. A graphical representation of this cohesive-frictional model is shown in Figure 2. All the other contact types between cells and with the drum walls present a purely frictional behaviour (see Fig. 3).

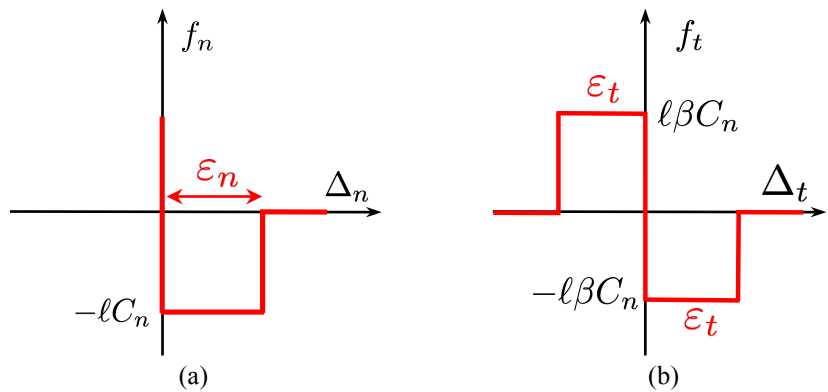

Figure 2. Behaviour of a cohesive interface a) Signorini relation in the normal direction; b) Coulomb friction law. $\Delta_n$ and $\Delta_t$ are the displacements between the two cells in the normal and tangential direction, respectively. $\varepsilon_n$ and $\varepsilon_t$ are related through $\varepsilon_n^2 + \varepsilon_t^2 = \varepsilon^2$ where $\varepsilon$ is the distance threshold.

The above model represents a rigid-plastic behaviour. No elastic strain is present because of the rigid nature of the elements. At a cohesive interface, reaching the stress threshold will trigger the plastic deformation (irreversible), that resembles a crack growth.

**Contact Dynamics**

The numerical simulations are carried out in the framework of a DEM method known as Contact Dynamics (CD) [9,10]. As in Molecular Dynamics (MD), in CD the equations of motion for all rigid bodies are integrated in time, taking into account the momentum conservation in the presence of the forces and moments due to contact between particles and, bulk forces such as gravity. In contrast to MD, an implicit scheme based on an iterative Gauss-Seidel algorithm is used in CD. This leads to unconditional stability of the time-stepping scheme that permits choosing larger time steps. The CD starts with a geometrical research of potential contacts in two steps. First, a rough selection of the neighbours with a search distance, followed by a narrower detection in which the positions of the geometrical features of the two particles potentially in contact are compared. Then, through an iterative algorithm, the forces and velocities are solved for all the potential contacts at the end of each time step. Finally, the particle velocities and rotations are calculated and the positions are updated. There is not a unique solution for perfectly rigid particles in CD but, the initialization of the contact forces at the beginning of the time step with the values found in the previous step reduces the set of solutions, such that the variation between solutions is below the numerical solution fluctuations [9].

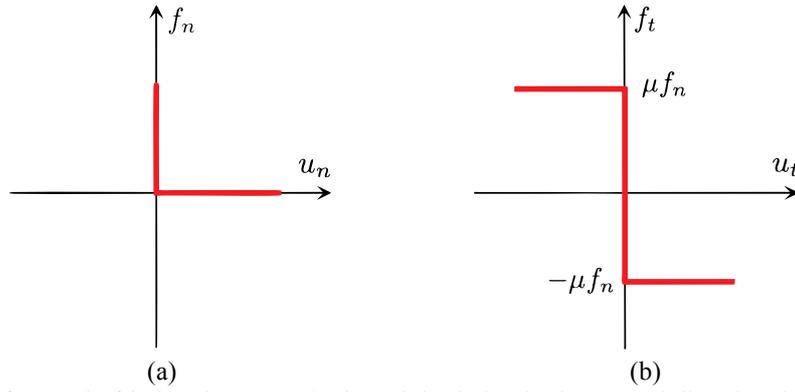

(a) (b)

Figure 3. Behaviour of a purely frictional contact a) Signorini relation in the normal direction, b) Coulomb friction law in the tangential direction. $u_n$ and $u_t$ denote the particles relative velocities in the normal and tangential directions, respectively.

Another parameter that must be set in CD is the restitution coefficient. Usually, it is defined from the relative velocities after and before a collision between two particles, such that it is zero for a completely inelastic shock and 1 for perfectly elastic collisions. The notion of restitution coefficient is included in CD by replacing the contact velocities ($u_n$ and $u_t$), defined in the Signorini and Coulomb relations at the end of each time step, by weighted means as follows:

$$u_n = \frac{u_n^+ + e_n u_n^-}{1+e_n}, u_t = \frac{u_t^+ + e_t u_t^-}{1+e_t}, \qquad Eq\ (1)$$

where $u_n^-$ and $u_t^-$ are the relative velocities at the beginning of the time step in the normal and tangential directions, respectively, $u_n^+$ and $u_t^+$ are the velocities at the end of the time step, and $e_n$ and $e_t$ are the restitution coefficients in the normal and tangential directions. The Signorini relation (see Fig. 3) determines that a contact takes place when $u_n = 0$ implying that $u_n^+ = -e_n u_n^-$, which is the classical definition of restitution coefficient.

**Sample setup and tests**

In order to simulate grinding inside a ball mill, a hollow cylinder of an internal diameter of 15 cm is filled with powder and balls. The powder volume is fixed as well as its material properties and those of the grinding bodies for all the studied cases (see Table 1). An important characteristic of our model is that all the elements have polygonal shapes, the powder particles are pentagons while the balls are hexadecagons. The use of polydisperse pentagons prevents from the creation of local crystallized structures often found in mono-disperse packings of hexagons and squares [11,12]. The powder particle sizes are defined using a uniform particle volume fraction from $d_{min}$= 0.002 m to $d_{max}$= 0.003 m. In general, the size of a polygonal particle is defined by the diameter of its circumscribed circle. Finally, the sample is tested by applying a constant speed of 50 rpm to the cylinder for a duration of 60 seconds. Snapshots at different times of a test show the evolution of the grinding process inside the drum (see Fig. 4). The flow regime inside rotating drums is generally described by the Froude number $Fr = \omega^2 R/g$ [13]. In our simulations, $Fr$ is 0.21 and the tested systems exhibit a cascading-cataracting regime.

Table 1. Material properties of the powder and balls used in all simulations.

| | |
|---|---|
| $C_n$ (MPa) | 1 |
| $\mu$ (-) | 0.4 |
| $\varepsilon$ (m) | 1.00E-06 |
| $\langle d_{cell} \rangle$ (m) | 5.00E-04 |
| $\rho_p$ (kg/m$^3$) | 2030 |
| $\rho_g$ (kg/m$^3$) | 11000 |

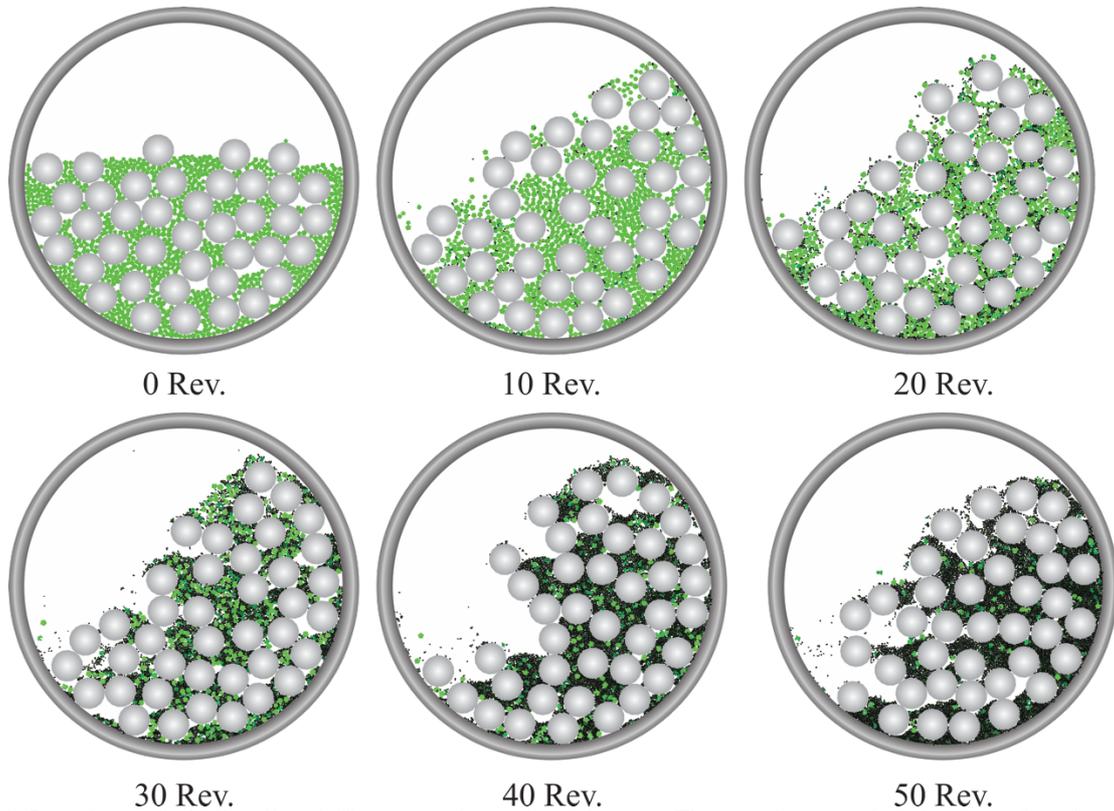

Figure 4. Snapshots of a ball mill at different numbers of revolutions. The powder particle colours go from bright green (intact) to black (highly damaged).

Two cases are considered in this study. In the first case, the effect of the ball size $D_b$ is studied, whereas in the second case the number of balls ($N_b$) is varied. In the first case, five samples were built with filling degree of 0.6 and a ratio between the balls and powder volumes ($V_b/V_p$) equal to 3.3. The filling degree is defined as the ratio of the apparent volume of the mixture powder-balls and the drum total volume. Since a constant amount of powder is used in all simulations and the ball size is variable, it is expected that the number of damaged powder particles will change from one sample to another.

The size $D_b$ is the same for all the balls in a given sample and takes values of 5, 10, 15, 20, 25 mm. Three samples of this case are displayed in Figure 5(a). For the second case, the size $D_b$ is constant and equal to 15 mm for the five sets of simulations. As in the first case, the powder volume is kept constant and thus the drums filled with different numbers $N_b$ of balls present different filling degrees different values of the ratio $V_b/V_p$ (see Table 2). In Figure 5(b) images of three samples of the second case are shown. In Table 2, some geometrical relations of the two studied cases are presented.

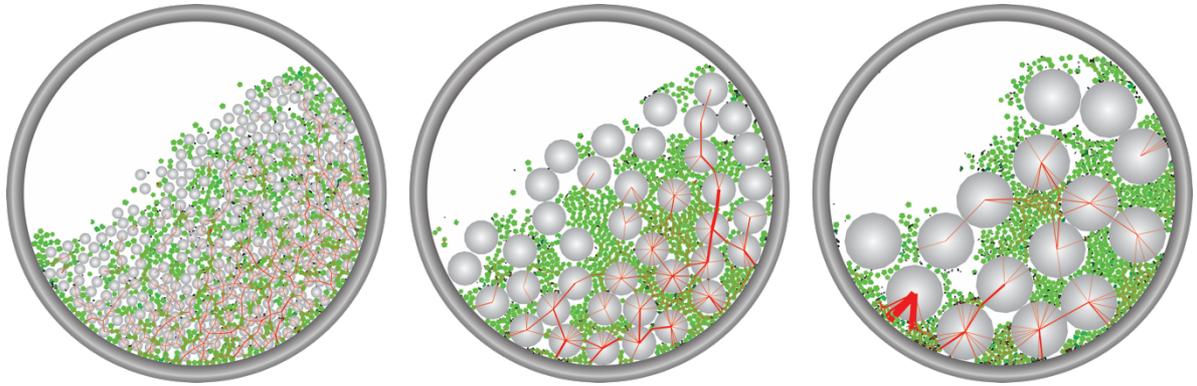

(a)

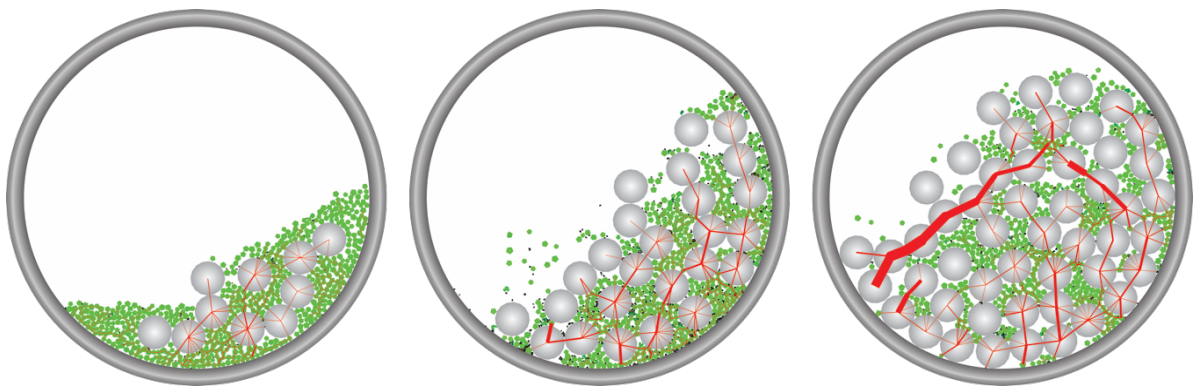

(b)

Figure 5. Snapshots of some of the samples tested. a) Samples with balls size $D_b$ of 5 mm, 15 mm and 25 mm from left to right; b) Samples with balls numbers $N_b$ of 10, 25 and 50. The thickness of the red lines is proportional to the magnitude of the normal force.

Table 2. Geometrical characteristics of the systems studied.

| First case | | | | Second case | | | |
|---|---|---|---|---|---|---|---|
| $D_b$ (mm) | $D_{drum}/D_b$ (-) | $D_b/\langle d_0 \rangle$ (-) | U (-) | $N_b$ (-) | Filling degree (-) | $V_b/V_p$ (-) | U (-) |
| 5 | 30 | 2 | 2.313 | 10 | 21.43% | 1.644 | 7.728 |
| 10 | 15 | 4 | 2.405 | 20 | 32.14% | 2.288 | 3.864 |
| 15 | 10 | 6 | 2.208 | 25 | 39.29% | 2.61 | 3.091 |
| 20 | 7.5 | 8 | 2.306 | 30 | 42.86% | 2.931 | 2.576 |
| 25 | 6 | 10 | 2.026 | 50 | 67.86% | 4.219 | 1.546 |

## 3. RESULTS

### Effect of ball size ($D_b$)

Figure 6(a) shows the mean powder particle size $\langle d \rangle$ normalized by the initial mean size $\langle d_0 \rangle$ as a function of the number of revolutions for different values of ball size $D_b$. The filling degree, total ball volume $V_b$, and total powder volume $V_p$ have the same value in all cases. This means that $D_b N_b$, where $N_b$ is the total number of balls is constant, and thus when we increase $D_b$, the number of balls declines. We observe slow size diminution at the beginning. Then, the size diminution is accelerated almost exponentially for the next 10 revolutions before slowing down again exponentially with powder particle size reaching a value close to the cell size. The transient occurs more or less early depending on the ball size, but we do not observe a monotonic dependence. Figure 6(b) displays the evolution of the total specific surface S normalized by the initial specific surface $S_0$ with the number of revolutions. The specific surface increases nonlinearly with the number of revolutions. Interestingly, besides $D_b$=5 and $D_b$=25, the evolution curves coincide for all other values of ball diameter. This is consistent with the data points of Figure 6(b) in which the evolution of the mean powder particle diameter for $D_b$=5 and $D_b$=25 is slower than for other diameters.

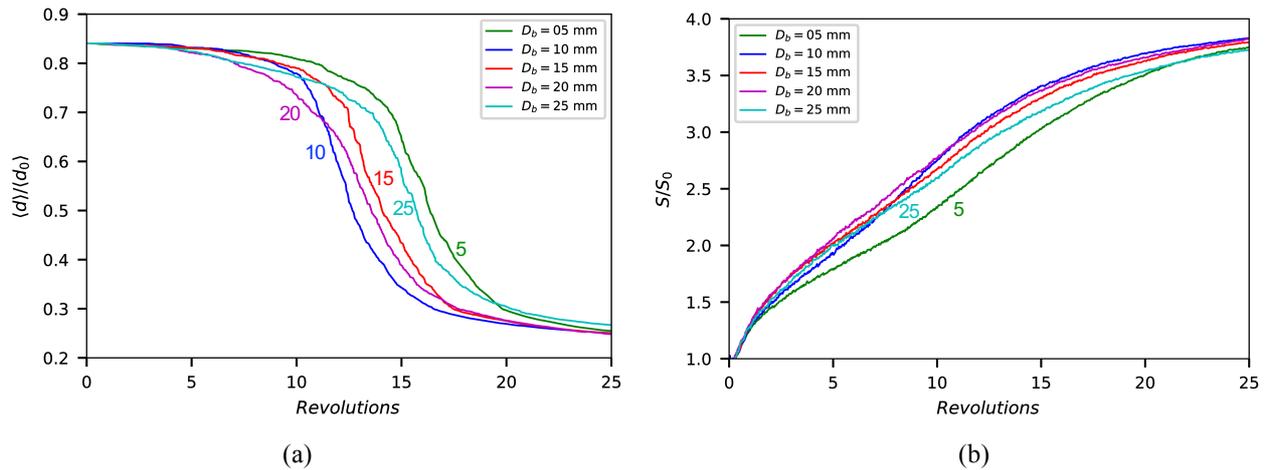

(a)                                            (b)

Figure 6. Evolution with the number of revolutions of a) normalised mean particle size, b) normalised specific surface, for different value of ball size. The filling degree, total balls volume $V_b$, and powder volume $V_p$ are constant. Each curve consists of 1000 data points.

In the case of the smallest value of $D_b$, where $D_b/D_p=2$, the milling is very similar to the case of powder ground without balls. Since the breakage events are concentrated on at the downstream of the free surface, as observed in Figure 7, the dominant breakage mechanism is the impact of particles, including both powder particles and balls, with the powder particles downstream of the free surface. Late grinding occurs in this case due to the balls low inertia: smaller amounts of kinetic energy are carried by small balls in comparison to big ones, and therefore the impact energy is transmitted to the powder in small amounts. As noted by Erdem and Ergün [3], the small balls are suitable for reducing the small powder particles rather than the big ones which are mainly broken by impacts of high collisional forces.

In the case of large values of $D_b$ with $D_b/D_p =10$, fewer impacts but of higher magnitude are expected [14]. In Figure 7 the multiple breakage events that are located at the downstream boundary with the drum wall can be linked to cases in which one or multiple grains are crushed between the wall and a ball that approaches with a large amount of kinetic energy. Also, this map shows that multiple breakage events occur in the space between balls (with its signature as dense rings), which is a feature not observed for $D_b$=5 mm. However, in this case, the grinding process is slower than for other diameters because the powder particles are trapped in the pores between the balls, becoming inaccessible and therefore protected. This result was also found in [3].

Between the two extreme cases discussed above, the intermediate cases show a gradual grinding transition that does not seem to depend on the ball size. This must be understood as the result of the fact the total volume of balls is kept constant. As the kinetic energy is proportional to the volume, the observed behaviour suggests that the surface created by milling is proportional to the kinetic energy and hence should be independent of ball size.

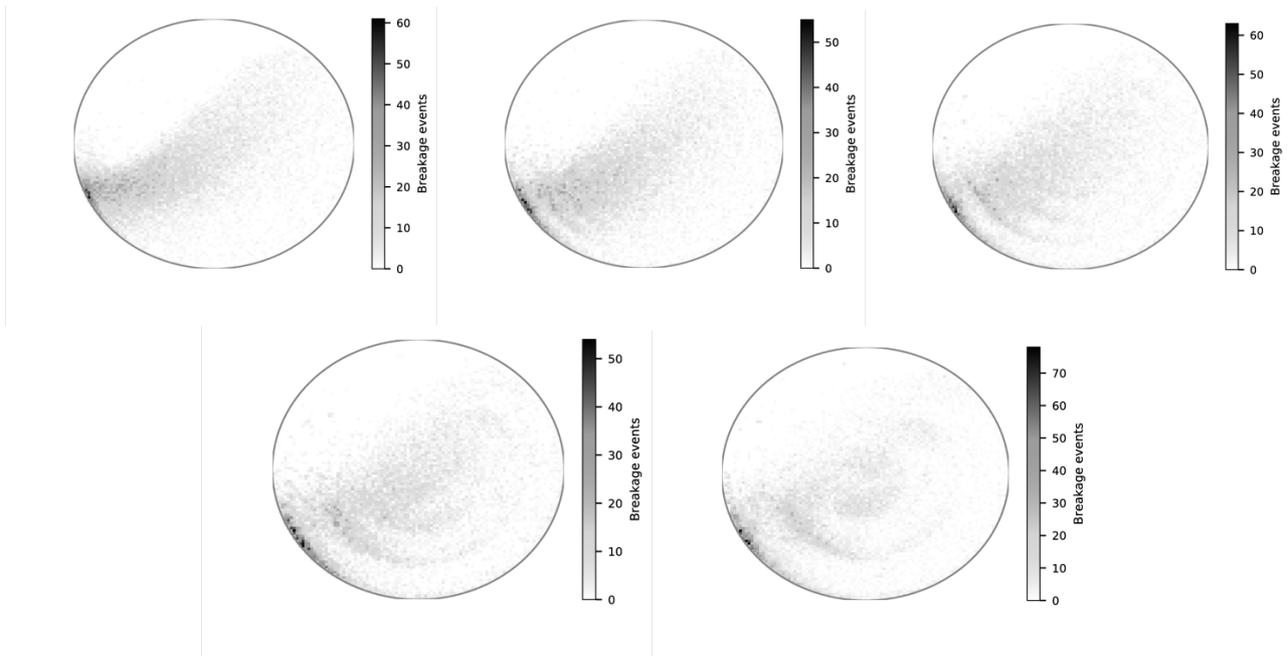

Figure 7. Spatial localization of breakage events in drums filled with balls of variable size: from left to right 5, 10, 15, 20, 25 mm, respectively.

**Effect of the number of balls ($N_b$)**

Figures 8(a) and 8(b) show the evolution of the mean powder particle size and specific surface with the number of revolutions for different values of the number $N_b$ of balls for constant values of ball size $D_b$ and total powder volume $V_p$. We see that the transition in terms of the specific surface and average powder particle size is increasingly faster as the number of balls increases except for $N_b$=50. As the total volume of the balls increases, the collisions between balls becomes increasingly dominant and more energy is dissipated as a result of ball collisions. In the limit $N_b$=50, this dissipation is excessively high. Indeed, as observed in a snapshot of the ball mill in Figure 9 in the case $N_b$=50 the balls often form long impulsive force chains. In the other extreme case of $N_b$=10, the filling degree is low, implying that the flow regime of this system is slumping-rolling, in which some of the particles can remain intact at the core, rather than the cascading-cataracting regime, characterized by the impact between powder and balls particles. For this reason, the transition curve in Figure 8 for $N_b$=10 is slightly different and the specific surface tends to a lower asymptotic value than for other values of $N_b$. Figure 10 shows maps of breakage events for the different number of balls tested. We observe that breakage occurs in all cases almost everywhere inside the flow with more events downstream of the flow.

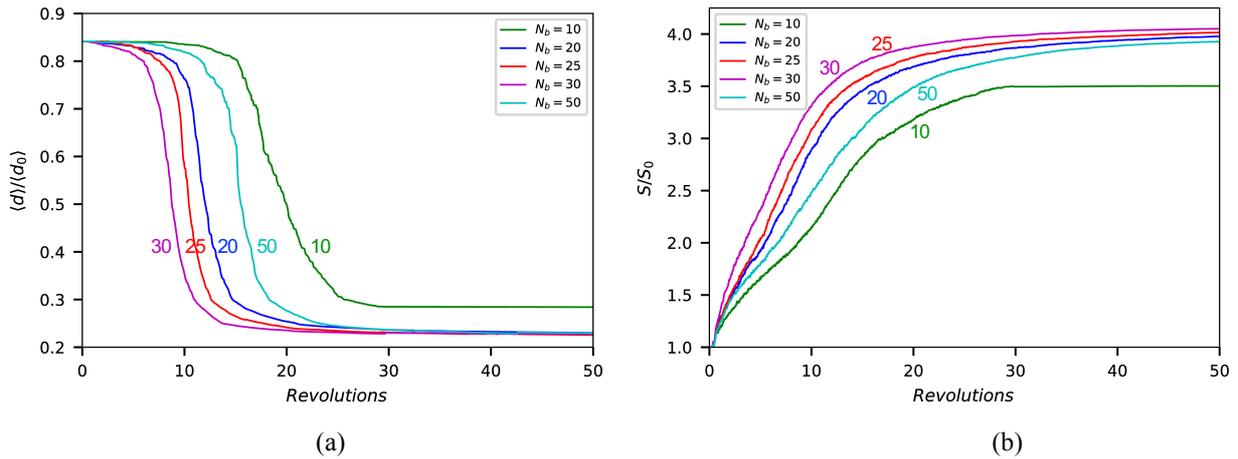

(a)                    (b)

Figure 8. Evolution with the number of revolutions of a) normalized mean particle size and b) normalised specific surface, for different numbers of balls ($N_b$). The ball size $D_b$ and the powder volume $V_p$ are constant. Each curve consists of 1000 data points.

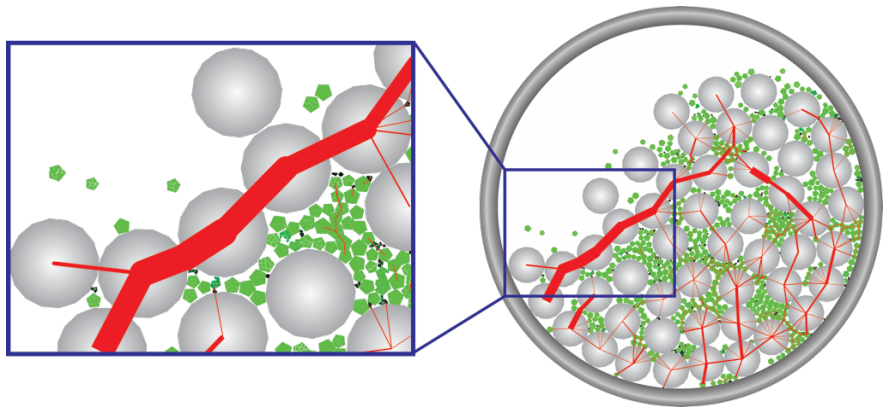

Figure 9. Force chains in a simulation with $N_b = 50$, the thickness of the red lines is proportional to the magnitude of the normal force.

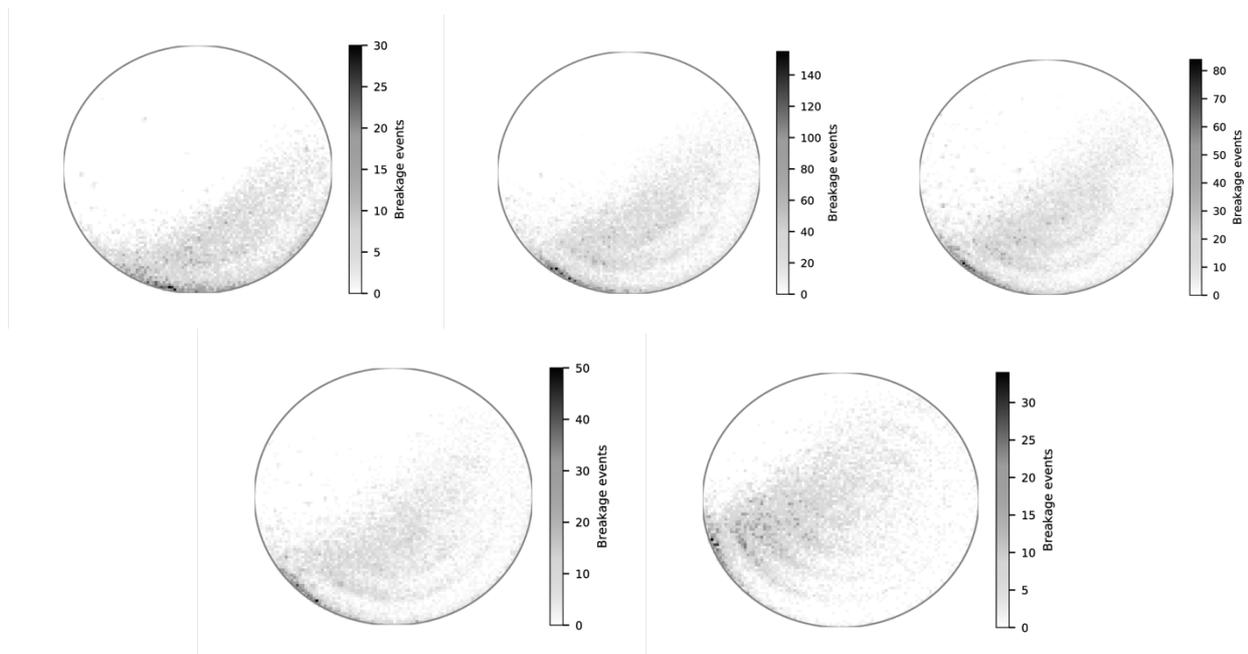

Figure 10. Spatial localization of breakage events in drums filled with balls of size 15 mm. From left to right the number of balls ($N_b$) is 10, 20, 25, 30, 50, respectively.

## 4. CONCLUSIONS

In this paper, we presented the bonded-cell method (BCM) developed on a Contact Dynamics platform and applied to the grinding of brittle materials inside ball mills. We performed simulations of a rotating cylinder containing a powder (breakable pentagonal particles) and balls (unbreakable disks). The effect on the reduction of particle size and changes of the specific surface were investigated for two groups of parameters. In the first group, the ball size was varied with a constant total volume of balls. In the second group, the number of balls was varied. In both cases the total number of powder particles was kept constant. We showed that the grinding behaviour (transient from the initial breakage to the ultimate state of nearly no breakage event) is mainly influenced by the number of balls. The grinding is faster as the number of balls and hence the total kinetic energy increases. However, for a large number of balls, this trend is counterbalanced by enhanced energy dissipation due to increased collisions between balls. In the case where the total volume of balls is kept constant, changing the ball size does not affect the evolution of grinding as the total kinetic energy is nearly the same. Here too, the extreme values of ball size correspond to special flow configurations that govern the grinding behaviour. Our simulations show that the BCM approach can be used efficiently for realistic simulations of ball mills with a large number of revolutions.

## ACKNOWLEDGMENTS

The authors would like to acknowledge the funding and expertise from the CEA.